\documentclass[apj]{emulateapj}

\newcommand{\etal}{et al.}

\newcommand\chandra{{\it Chandra}}

\newcommand\xmm{{\it XMM-Newton}}

\def\nh{\hbox{$N_{\rm H}$}}

\def\snr{Kes~79}
\def\psr{\rm{PSR J1852$+$0040}}
\def\src{\rm{CXOU~J185238.6$+$004020}}
\def\one{\rm{1E~1207.4$-$5209}}

\def\simlt{\mathrel{\hbox{\rlap{\hbox{\lower4pt\hbox{$\sim$}}}\hbox{$<$}}}}
\def\simgt{\mathrel{\hbox{\rlap{\hbox{\lower4pt\hbox{$\sim$}}}\hbox{$>$}}}}

\slugcomment{Received 2007 April 10; accepted 2007 May 7}

\shorttitle{X-ray Timing of PSR J1852+0040}
\shortauthors{Halpern et al.}

\begin{document}

\title{X-ray Timing of PSR J1852+0040 in Kesteven~79: \\
Evidence of Neutron Stars Weakly Magnetized at Birth}

\author{J. P. Halpern\altaffilmark{1}, E. V. Gotthelf\altaffilmark{1},
F. Camilo\altaffilmark{1}, and F. D. Seward\altaffilmark{2}}

\altaffiltext{1}{Columbia Astrophysics Laboratory, Columbia University,
New York, NY 10027}
\altaffiltext{2}{Smithsonian Astrophysical Observatory, Cambridge, MA 02138}

\begin{abstract}
The 105-ms X-ray pulsar J1852+0040 is the central
compact object (CCO) in supernova remnant \snr.  We report a
sensitive upper limit on its radio flux density of $12\,\mu$Jy at
2~GHz using the NRAO Green Bank Telescope.
Timing using the {\it Newton X-Ray Multi-Mirror Mission} (\xmm)
and the \chandra\ {\it X-ray Observatory} over a 2.4~yr span
reveals no significant change in its spin period.
The $2\sigma$ upper limit on the period derivative leads,
in the dipole spin-down formalism, to an energy loss rate
$\dot E < 7 \times 10^{33}$~ergs~s$^{-1}$,
surface magnetic field strength $B_p < 1.5 \times 10^{11}$~G,
and characteristic age $\tau_c \equiv P/2\dot P> 8$~Myr.
This value of $\tau_c$
exceeds the age of the SNR by 3 orders of magnitude,
implying that the pulsar was born spinning
at its current period.  However, the X-ray luminosity of \psr,
$L_{\rm bol} \approx 3 \times 10^{33}\,(d/7.1\ {\rm kpc})^2$ ergs~s$^{-1}$,
is a large fraction of $\dot E$, which challenges the
rotation-powered assumption.  Instead, 
its high blackbody temperature $kT_{\rm BB} = 0.46\pm 0.04$ keV,
small blackbody radius $R_{\rm BB} \approx 0.8$~km, and large pulsed fraction
$f_p \approx 80\%$, may be evidence of accretion onto a polar cap, possibly
from a fallback disk made of supernova debris.
If $B_p < 10^{10}$~G, an accretion disk can
penetrate the light cylinder and interact with the magnetosphere
while resulting torques on the neutron star remain within the
observed limits. A weak $B$-field is also inferred in another CCO,
the 424-ms pulsar \one, from its steady
spin and soft X-ray absorption lines.
We propose this origin of radio-quiet CCOs:  
the magnetic field, derived from a turbulent
dynamo, is weaker if the NS is formed spinning slowly,
which enables it to accrete SN debris.  Accretion excludes
neutron stars born with both $B_p < 10^{11}$~G 
and $P > 0.1$~s from radio pulsar surveys,
where $B_p < 10^{11}$~G is not encountered except
among very old ($\tau_c > 40$~Myr) or recycled pulsars.
Finally, such a CCO, if born in SN~1987A,
could explain the non-detection of a pulsar there.
\end{abstract}


\section {Introduction}

A compact X-ray source, \src,
was found in the center of the supernova remnant
(SNR) Kes~79 by \cite{sew03}.  Using \xmm\ in 2004 October,
we discovered 105-ms X-ray pulsations from this CCO \citep[Paper 1]{got05},
also named \psr.  Its pulsed fraction was 
as high as 86\% and it had an apparently thermal X-ray spectrum.
From the discovery observations, only an upper limit on its period
derivative could be determined, $\dot P < 7 \times 10^{-14}$~s~s$^{-1}$,
which left the energetics of the NS and the mechanism of its
X-ray emission ambiguous.  The two most plausible models explored in Paper 1,
a rotation-powered pulsar, and accretion from fallback material,
each encountered some difficulties.

Here, we present three new observations,
two from \xmm\ and one from \chandra, that over a 2.4 year span refine
the spectral and timing properties of \psr.  In addition, we report
the negative result of a search for radio pulsations in a deep GBT pointing.
From the assembled X-ray timing of \psr, a sensitive new upper limit is
obtained on its period derivative, suggesting that the underlying cause
of its unusual properties is a weak magnetic field,
$B_p < 1.5 \times 10^{11}$~G.
Unlike canonical radio pulsars with $B_p \sim 10^{12-13}$~G, a young NS
that is weakly magnetized and spinning slowly at birth is able
to accrete from material that is thought to remain following a
supernova explosion. We discuss this as a model for the properties
of radio-quiet CCOs, including the similar pulsar \one.

\section{\xmm\ and \chandra\ Observations}

The two observations previously reported in Paper~1 were obtained with
\xmm\ on 2004 October 18 and 23.  Two more observations, reported
here, were made on 2006 October 8 and 2007 March 20
using the same instrument modes, and exposure times of 30~ks each.
The pn CCD of the European Photon Imaging Camera (EPIC;
\citealt{tur03}), was operated in ``small window'' mode
with a $4\farcm3 \times 4\farcm3$ field-of-view (FOV) and 5.7~ms time
resolution to resolve the pulsations of \psr.
The EPIC~MOS CCDs were exposed in
``full frame'' mode with a $30^{\prime}$ diameter FOV and time
resolution of 2.7~s. For all three cameras the medium
density filter was used.  We processed all four observations
with Science Analysis System
version xmmsas\_20060628\_1801-7.0.0, using the same techniques as in Paper~1.

\begin{figure}[t]
\centerline{
\hfill
\includegraphics[width=0.95\linewidth,angle=270]{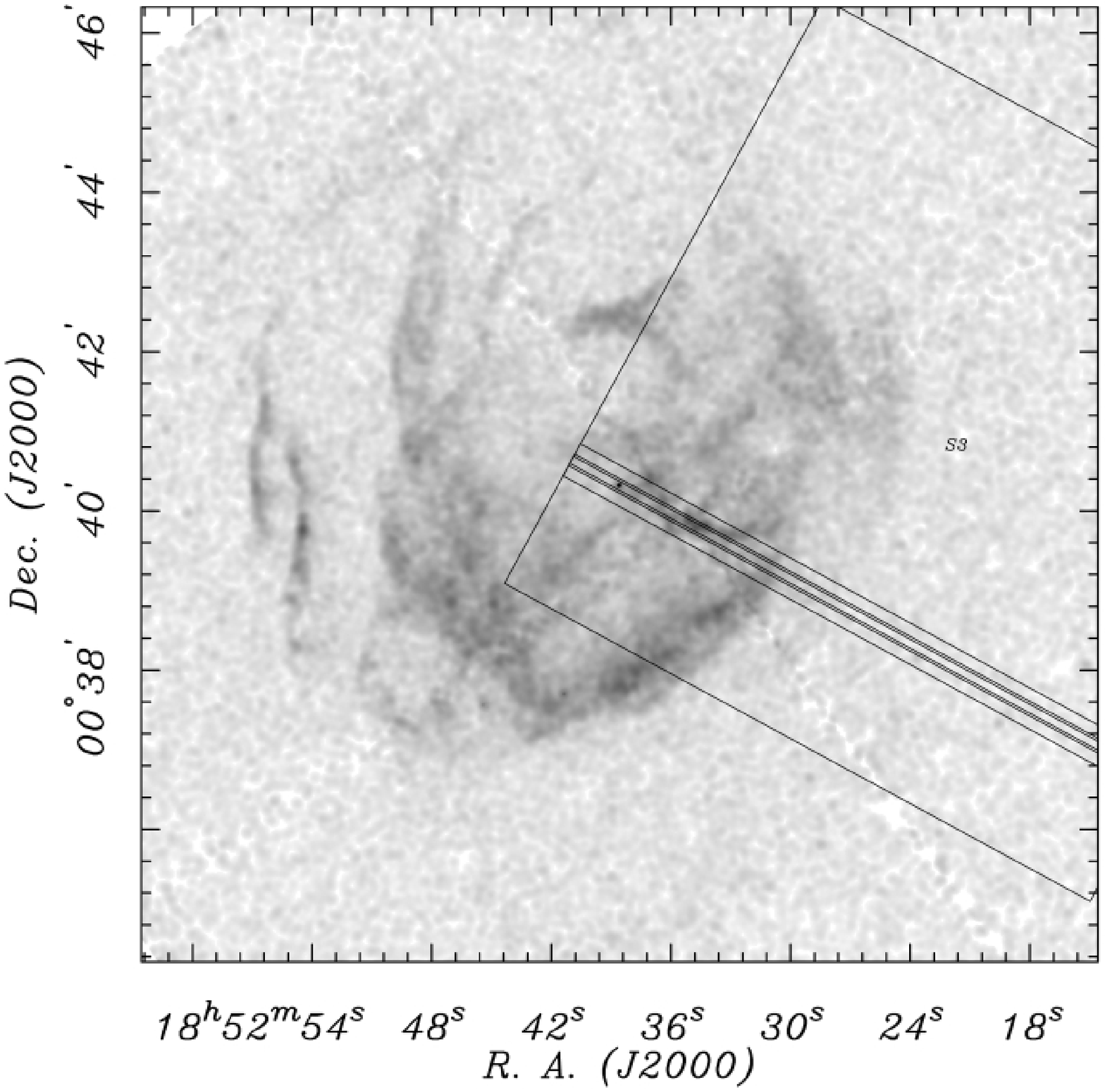}
\hfill
}
\caption{Geometry of the 2006 November 23
\chandra\ CC-mode observation of \psr,
showing the outline of the S3 CCD and the columns used for source and
background extraction, overlaid on the 2001 July 31
ACIS-I observation of \citet{sew03}.
(This broadband image of \snr\ is square-root
scaled to emphasize faint emission features and background.)
The counts were summed along the length of the
columns, oriented at position angle $241^{\circ}\!.5$
east of north. Both spectral and timing data for the pulsar were
extracted from the six central columns. The
background spectrum was obtained from nine neighboring columns on either
side ({\it flanking lines}), separated from the source by a
two-column gap.
}
\label{image}
\end{figure}

A 32~ks \chandra\ observation was performed on 2006 November~23
using the Advanced Camera for Imaging and Spectroscopy (ACIS)
operated in continuous-clocking (CC) mode to provide a time resolution
of 2.85~ms.  The back-illuminated ACIS-S3 CCD was used, and
the Scientific Instrument Module (SIM) was offset in the $-Z$
direction to place the pulsar $0^{\prime}\!.85$ from the edge of the
CCD to reduce contamination from the thermal remnant. To achieve
the fast timing in CC mode, one spatial dimension of the CCD image is
sacrificed by integrating along the column direction during each CCD
readout cycle. The geometry of the observation is illustrated in
Figure~\ref{image}, overlaid on an image of \snr\ previously
obtained with ACIS-I in timed exposure mode on 2001 July 31 \citep{sew03}.
We also reanalyzed the 2001 observation for the spectral properties
of the pulsar.  During both observations the CCD
background count rate was stable and no time filtering was necessary.
All photon arrival times were adjusted by the standard processing to
account for the spacecraft dithering, and SIM offset in the case of
the CC-mode observation.  Reduction and
analysis of the \chandra\ data are all based on the standard software
package CIAO (v3.4) and CALDB (v3.3).

\subsection{Timing Analysis}

\begin{deluxetable*}{llccccc}[h]
\tablecolumns{7}
\tablewidth{0pt}
\tablecaption{X-ray Timing of \psr}
\tablehead{
\colhead{Mission} & \colhead{Date} & \colhead{Epoch} & \colhead{Duration} &
\colhead{Rate\tablenotemark{a}} & \colhead{Period\tablenotemark{b}} &
\colhead{$f_p$\tablenotemark{c}} \\
& \colhead{(UT)} & \colhead{(MJD)} & \colhead{(s)} & \colhead{(s$^{-1}$)}
& \colhead{(ms)}  & \colhead{(\%)}\\
}
\startdata
{\it XMM} & 2004 Oct 18 & 53,296.001 & 30,587 & 0.041(2) & 104.912643(18) & 86(16)\\
{\it XMM} & 2004 Oct 23 & 53,301.985 & 30,515 & 0.046(2) & 104.912600(27) & 61(16)\\
{\it XMM} & 2006 Oct 08 & 54,016.245 & 30,243 & 0.045(2) & 104.912593(20) & 77(15)\\
\chandra\ & 2006 Nov 23 & 54,062.257 & 32,165 & 0.022(2) & 104.912612(19) & 100(16)\\
{\it XMM} & 2007 Mar 20 & 54,179.878 & 30,506 & 0.043(2) & 104.912609(19) & 71(15)\\
\enddata
\tablenotetext{a}{\footnotesize For \xmm, background and dead-time corrected
count rate in a $15^{\prime\prime}$ radius aperture.
Statistical ($\sqrt{N}$) uncertainty
in the last digit is given in parentheses.}
\tablenotetext{b}{\footnotesize Period derived from a $Z^2_1$ test. Period
uncertainty is $1\sigma$ computed by the Monte Carlo method described in
\citet{got99}.}
\tablenotetext{c}{\footnotesize Pulsed fraction defined as
$f_p \equiv N({\rm pulsed})/N({\rm total})$. Unpulsed level set at lowest
point of 20-bin folded lightcurve.}
\label{timetable}
\end{deluxetable*}

\begin{deluxetable*}{lcccccc}
\tablecolumns{7}
\tablewidth{0pt}
\tablecaption{X-ray Spectral Fits for \psr \label{spectable}}
\tablehead{
\colhead{}  & \colhead{2001 Jul 31} & \colhead{2004 Oct 18} & \colhead{2004 Oct 23} & \colhead{2006 Oct 08} &
\colhead{2006 Nov23} & \colhead{2007 Mar 20}\\
\colhead{Parameter}  & \colhead{\chandra} & \colhead{{\it XMM}} & \colhead{{\it XMM}} & \colhead{{\it XMM}} & \colhead{\chandra} &
\colhead{{\it XMM}}
}
\startdata
\multispan7{\hfill \hbox{Blackbody~Model}\hfill \vspace{5pt}}\\
\tableline
$N_{\rm H}$ ($10^{22}$ cm$^{-2}$)                & $1.1 \pm 0.3$        & $1.4 \pm 0.3$        & $1.4 \pm 0.3$        & $1.4\pm0.3$         & $1.6\pm 0.3$
  & $1.3\pm 0.3$        \\
$kT_{\rm BB}$ (keV)                              & $0.50 \pm 0.04$      & $0.44\pm 0.04$       & $0.45\pm 0.03$       & $0.46\pm0.04$       & $0.41\pm 0.04$
  & $0.46\pm 0.04$      \\
$R_{\rm BB}$ (km)                                & $0.62\pm0.05$        & $0.92 \pm 0.09$      & $0.79 \pm 0.09$      & $0.74\pm0.08$       & $0.96\pm0.04$
  & $0.69\pm0.09 $      \\
$A_{\rm BB}$ ($10^{10}$~cm$^2$)                  & $4.8\pm0.7$          & $10.7 \pm 2.1$       & $7.8 \pm 1.7$        & $6.8 \pm 1.5$       & $11.5\pm4.4$
  & $6.0\pm1.5$         \\
$F_{\rm BB} (10^{-13}$~cgs)\tablenotemark{a}   & $2.0\pm0.3$          & $1.9 \pm 0.3$        & $2.0 \pm 0.3$        & $2.0\pm 0.3$        & $1.7\pm0.4$
  & $1.9\pm0.3$         \\
$L_{\rm BB, bol} (10^{33}$~cgs)\tablenotemark{b}& $3.2\pm 0.7$         & $4.3 \pm 1.1$        & $3.3 \pm 1.0$        & $3.2\pm 1.0$        & $3.5\pm1.7$
  & $2.9\pm1.1$         \\
$\chi^2_{\nu}({\nu})$                            & 0.71(18)             & 0.85(68)             & 0.67(70)             & 0.86(69)            & 0.88(22)
  & 0.67(69)            \\
\cutinhead{Power-law~Model} 
$N_{\rm H}$ ($10^{22}$ cm$^{-2}$)                & $2.6^{+0.7}_{-0.4}$  & $3.2^{+0.6}_{-0.6}$  & $3.1^{+0.6}_{-0.5}$  & $3.2^{+0.6}_{-0.5}$ & $3.4^{+0.8}_{-0.6}$  & $3.2^{+0.7}_{-0.6}$ \\
$\Gamma$                                         & $4.1^{+0.7}_{-0.4}$  & $4.9^{+0.7}_{-0.6}$  & $4.9^{+0.6}_{-0.5}$  & $4.8^{+0.6}_{-0.5}$ & $5.3^{+0.8}_{-0.7}$  & $4.9^{+0.6}_{-0.6}$ \\
$F_{\rm PL} (10^{-13}$~cgs)\tablenotemark{a}   & $2.0\pm 0.6$         & $1.9 \pm 0.43$       & $2.0 \pm 0.4$        & $2.1\pm0.5$         & $1.8\pm0.4$
  & $1.9\pm0.4$         \\
$L_{\rm PL} (10^{33}$~cgs)\tablenotemark{c}    & $1.2\pm 0.4$         & $1.1 \pm 0.2$        & $1.2 \pm 0.2$        & $1.2\pm0.3 $        & $1.1\pm0.3$
  & $1.2\pm0.3$         \\
$\chi^2_{\nu}({\nu})$                            & 0.94(18)             & 0.82(68)            & 0.68(70)             & 0.78(69)            & 0.81(22)
 & 0.71(69)            \\
\enddata
\tablecomments{\footnotesize Uncertainties are 68\% confidence intervals for two interesting parameters. Luminosities are computed for $d=7.1$~kpc.}
\tablenotetext{a}{\footnotesize Absorbed flux in the $1-5$ keV band in units of ergs~cm$^{-2}$~s$^{-1}$.}
\tablenotetext{b}{\footnotesize Unabsorbed, bolometric blackbody luminosity in units of ergs~s$^{-1}$.}
\tablenotetext{c}{\footnotesize Absorbed luminosity in the $1-5$ keV band in units of ergs~s$^{-1}$. This corrects an error in Table~2 of Paper~1.}
\end{deluxetable*}

\begin{figure}[t]
\centerline{
\hfill
\includegraphics[width=0.75\textwidth,angle=270]{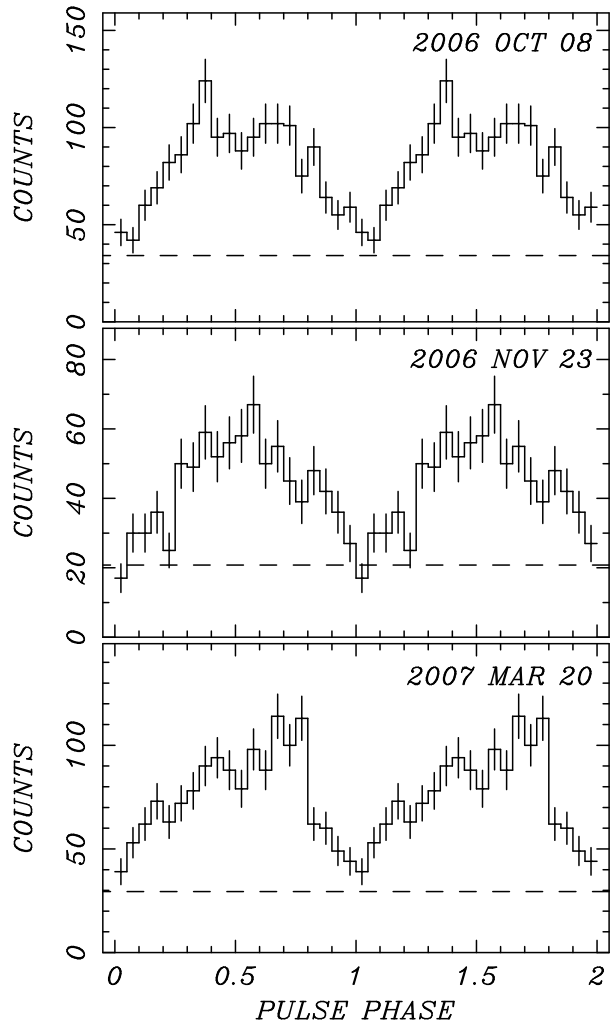}
\hfill
}
\vspace{0.3cm}
\caption{Folded light curves of \psr\ in the $1-5$ keV band using 
\xmm\ EPIC~pn on 2006 October 8 ({\it top}),
\chandra\ ACIS-S3 in CC mode on  2006 November 23 ({\it middle}),
and \xmm\ EPIC~pn on 2007 March 20  ({\it bottom}).
The estimated background level is indicated by the dashed line.
Phase zero is arbitrary.}
\label{timeplot}
\end{figure}

Photon arrival times were converted to the solar system
barycenter using the \chandra\ derived source coordinates
from Paper~1, R.A. = $18^{\rm h}52^{\rm m}38\fs57$,
decl. =  $+00\arcdeg40\arcmin19\farcs8$ (J2000.0),
and signal strength was maximized by selecting the $1-5$~keV energy
band.  From the \xmm\ observation of 2006 October 8,
a $Z^2_1$ test \citep{buc83} localizes the period to
$P = 104.912593(20)$~ms with a peak statistic of $Z^2_1 = 88$.
The resulting light curve, which has pulsed fraction
$f_p = 77\% \pm 15\%$ after accounting for SNR background,
is shown in Figure~\ref{timeplot}.  Here, we define the pulsed fraction as 
$f_p \equiv N({\rm pulsed})/N({\rm total})$,
where we choose the minimum of the folded light curve as
the unpulsed level.  The observation on 2007 March 20 yielded
consistent parameters, $P = 104.912609(19)$~ms with $Z^2_1 = 83$
and $f_p = 71\% \pm 15\%$.  These new values of
$f_p$ are intermediate between
the 2004 \xmm\ observations, which yielded
$f_p = 86\% \pm 16\%$ and $61\% \pm 16\%$, respectively,
weakening the marginal evidence of variability of
$f_p$ that was noted in Paper~1.
Similar analysis of the \chandra\ observation
of 2006 Novemeber 23 yields a formally identical period,
$P = 104.912612(19)$~ms with $Z^2_1 = 94$, and $f_p$ consistent
with 100\%.  The pulsed fraction is less reliable in CC-mode data
due to systematics of estimating the contamination from SNR background.
Table~\ref{timetable} is a summary of all of the available
timing data on \psr.

While the new observations extend the sampled
time span by a factor of 150 since Paper~1,
there is still no significant detection of a period derivative.
A $\chi^2$ fit to the five measurements of $P$ yields the
formally negative $<\dot P> = (-3.4 \pm 2.7) \times 10^{-16}$~s~s$^{-1}$
that is, however, consistent with zero at the $\approx 1\sigma$ level.
We are unable to link any of the observations with a
definite cycle count in order to improve on this measurement.
If we adopt the $2\sigma$ upper limit $\dot P < 2.0 \times 10^{-16}$,
the dipole spin-down formalism for an isolated pulsar implies
an energy loss rate $\dot E = -I\Omega\dot\Omega = 
4\pi^2I\dot P/P^3  < 7 \times 10^{33}$~ergs~s$^{-1}$,
surface magnetic field strength
$B_p = 3.2\times10^{19}\sqrt{P\dot P} < 1.5 \times 10^{11}$~G,
and characteristic age $\tau_c \equiv P/2\dot P > 8$~Myr.
These derived quantities pose problems for a rotation-powered
X-ray source, as discussed in \S 4.1.

\subsection{Spectral Analysis}

\begin{figure}[t]
\centerline{
\hfill
\includegraphics[width=0.34\textwidth,angle=270]{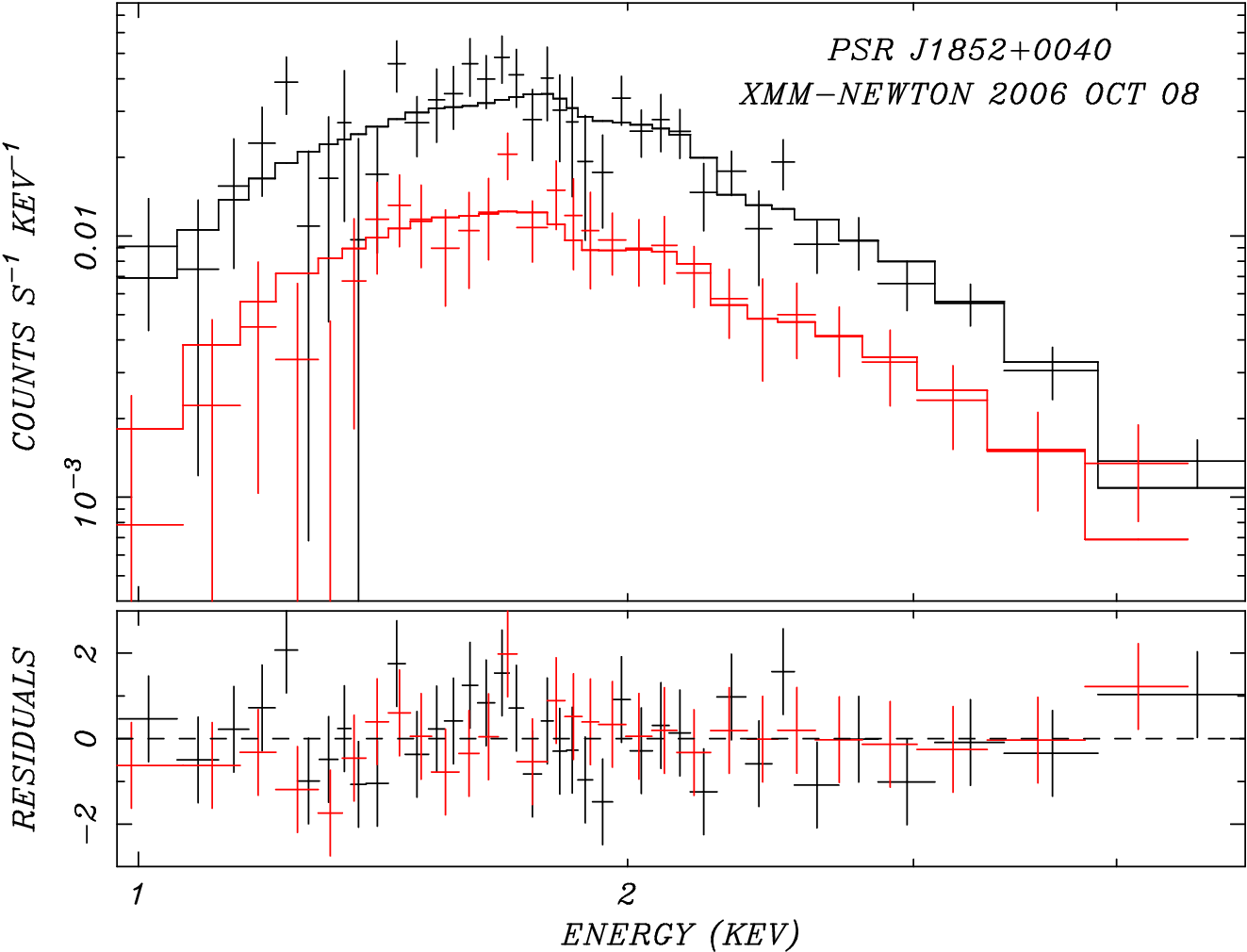}
\hfill
}
\vspace{0.2cm}
\centerline{
\hfill
\includegraphics[width=0.34\textwidth,angle=270]{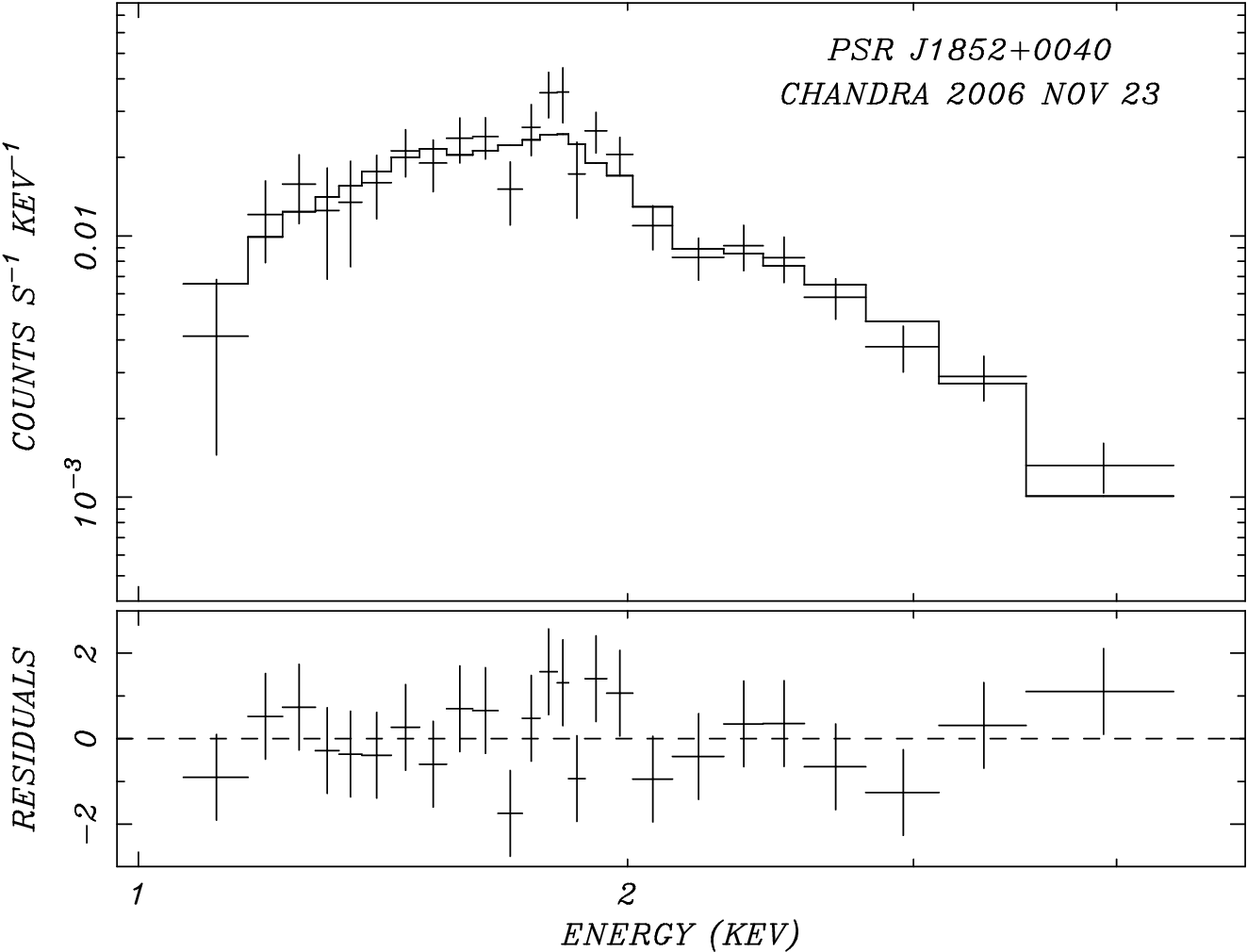}
\hfill
}
\vspace{0.2cm}
\centerline{
\hfill
\includegraphics[width=0.34\textwidth,angle=270]{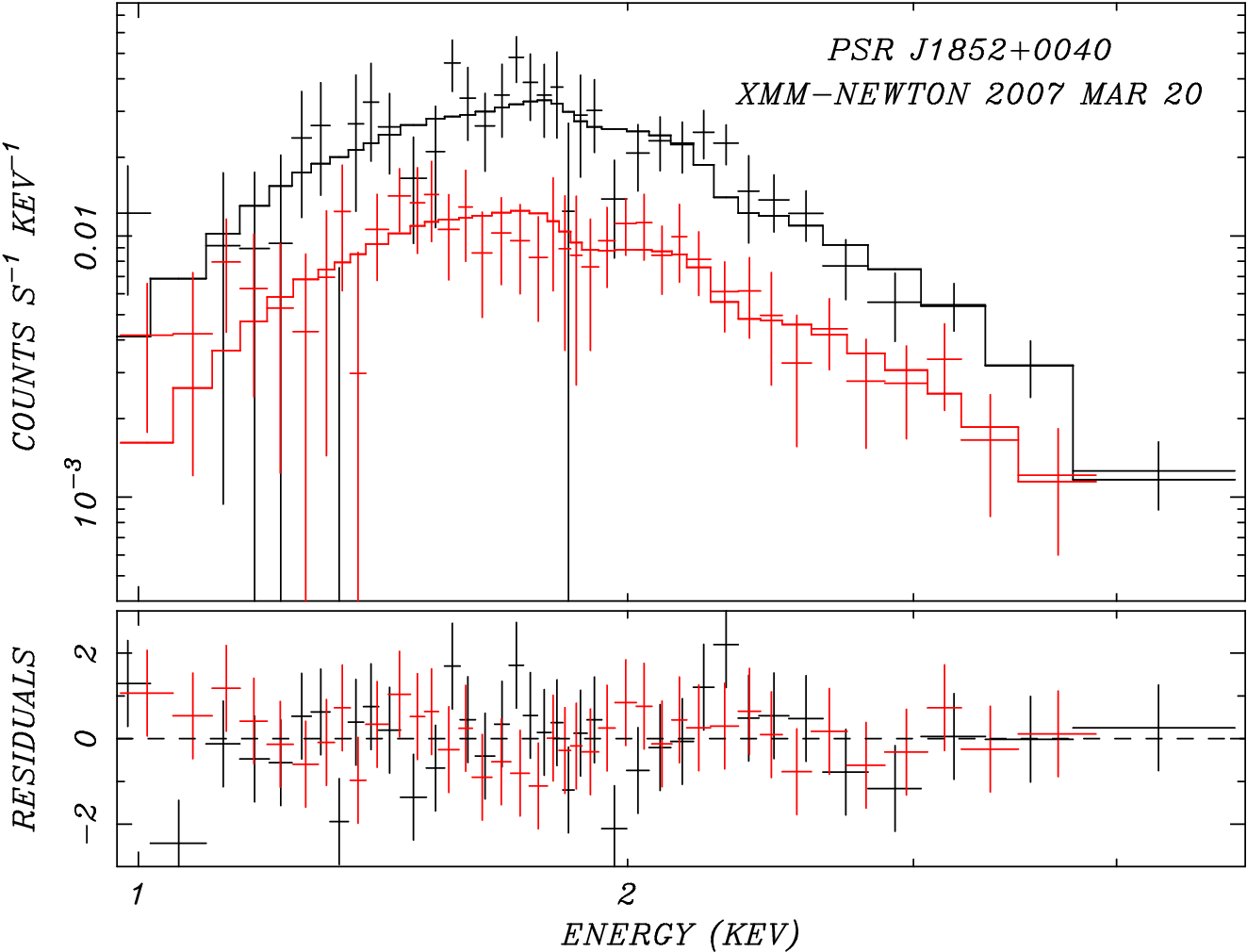}
\hfill
}
\caption{X-ray spectra of \psr\ in Kes~79.
{\it Top}: The \xmm\ spectrum on 2006 October 8. Data from the EPIC~pn
({\it black, upper}) and EPIC MOS ({\it red, lower}) with
the best-fit blackbody model using parameters in Table~\ref{spectable}.
The residuals from the fits are in units of $\sigma$.
{\it Middle}: Spectrum on 2006 November 23 from the \chandra\ ACIS-S3
in CC mode. 
{\it Bottom}:  The \xmm\ spectrum on 2007 March 20.
[{\it See the electronic version of the Journal for a color
version of this figure.}]}
\label{specfig}
\end{figure}

Following the analysis method in Paper~1 for each \xmm\ observation,
spectra from the two EPIC MOS cameras (MOS1, MOS2)
were co-added.  The EPIC~MOS and EPIC~pn spectra
were then fitted simultaneously using the {\tt XSPEC} (v12.21)
package.  The results of spectal
fits using either a blackbody or an absorbed power-law model
are presented in Table~\ref{spectable}.  An acceptable
$\chi^2$ statistic is obtained using either model.
However, the blackbody is preferred over
the power law based on its derived column density of $\nh =
(1.4\pm0.3) \times 10^{22}$~cm$^{-2}$, which is consistent with that
found for the SNR \citep[$\nh \approx 1.6 \times
10^{22}$~cm$^{-2}$,][]{sun04}. The fitted column density for the
power-law model, $\nh = 3.2^{+0.6}_{-0.5} \times 10^{22}$~cm$^{-2}$,
is significantly larger than the integrated 21 cm Galactic value of
$2\times 10^{22}$~cm$^{-2}$ in this direction \citep{dic90}.
The best-fit blackbody
model yields temperature $kT_{\rm BB} = 0.46\pm0.04$~keV (see
Fig.~\ref{specfig}).  Adopting a distance $d=7.1$~kpc from \cite{cas98},
the bolometric blackbody luminosity
$L_{\rm BB, bol} \approx 3.0\times10^{33}$~ergs~s$^{-1}$,
corresponding to blackbody area 
$A_{\rm BB} \approx 7 \times 10^{10}\,(d/7.1\ {\rm kpc})^2$~cm$^2$
or $\approx 0.4\%$ of the NS surface.
These results are consistent with the spectral parameters from the previous
\xmm\ observations in 2004 October, described in Paper~1 and summarized
here in Table~\ref{spectable}, indicating that
the flux and spectrum have remained steady. 

The \chandra\ spectra from the 2001 and 2006 observations
were each prepared by grouping a minimum of 40 counts
per channel, and fitted using the {\tt XSPEC} software.  The
standard spectral response matrices were generated for the target
location on the CCD and all spectra were corrected for the effects of
charge-transfer inefficiency; however, the spectral gain is not
calibrated well in CC-mode.   The resulting spectral parameters from
the \chandra\ observations, listed in Table~\ref{spectable},
are similar to the ones from \xmm.
The blackbody fits to the new \xmm\ and \chandra\ spectra
are shown in Figure~\ref{specfig}.

\section{Search for a Radio Pulsar}

On 2002 October 29 we observed the location of \src\ with the ATNF Parkes
telescope in NSW, Australia.  At that time the period of the pulsar was not
yet known, so we employed standard pulsar searching techniques.
We used the central beam of the multibeam
receiver at a frequency of 1374\,MHz, with 96 channels across a bandwidth
of 288\,MHz in each of two polarizations.  The integration time was
6.8\,hr, during which total-power samples were obtained every 0.73\,ms
and recorded for off-line analysis.  We searched the
dispersion measure range 0--2500\,cm$^{-3}$\,pc (twice the maximum
Galactic DM predicted for this line of sight by the Cordes \& Lazio
2000 electron density model; for the distance of 7.1\,kpc,
the predicted DM is 440\,cm$^{-3}$\,pc).  The search
followed closely that described in more detail in \citet{cam06}.
No promising candidate periods were identified.

Following the X-ray discovery of the pulsar period, we made a deeper search
using the GBT.  We did this on 2005 June 10,
observing the pulsar for 11.1\,hr with the Spigot spectrometer \citep{kap05},
sampling 768 frequency channels across a bandwidth of 600\,MHz
centered on 1950\,MHz every 81.92\,$\mu$s.  In addition to performing
a standard blind search, we also folded the data at a number of DMs
for periods close to the X-ray value
to increase the sensitivity, but again found no candidates.
We used the PRESTO software package to analyze these data \citep{ran01,ran02}.

Applying the standard modification to the radiometer equation, assuming a
quasi-sinusoidal pulse shape
(as in X-rays; see Fig.~\ref{timeplot}), and accounting
for a sky temperature at this location of 5\,K, we were sensitive to a
pulsar with period 105\,ms having a period-averaged flux density at 2\,GHz
$\ga 12\,\mu$Jy.  Converting this to the more common 1.4\,GHz pulsar search
frequency, using a typical spectral index of --1.6 \citep{lor95},
results in $S_{1.4} \la 0.02$\,mJy.  For a distance of $\approx 7$\,kpc,
this corresponds to a pseudo-luminosity limit of $L_{1.4} \equiv S_{1.4}
d^2 \la 1$\,mJy\,kpc$^2$.  Only one young radio pulsar has a smaller
radio luminosity than this \citep[the pulsar in 3C58;][]{cam02}.

Based on these results, it is therefore somewhat unlikely that
PSR~J1852+0040 is presently a radio pulsar beaming toward the Earth.
More likely, it is either beaming away \citep[observed radio pulsars with
similar periods, or ages $\la 10$\,kyr, have beaming fractions of $\sim
0.3$; see][]{tau98}, or not an active radio pulsar at all.

\clearpage
\section{Discussion}

\subsection{Rotation-Powered Pulsar?}

Potential obstacles to the interpretation
of \psr\ as a rotation-powered pulsar were identified
in Paper~1.  The newly reduced upper limit on
its $\dot P$, by more than 2 orders of magnitude,
turns those difficulties into major objections.
First, the X-ray luminosity of \psr,
$L_{\rm bol} \approx 3 \times 10^{33}\,(d/7.1\ {\rm kpc})^2$ ergs~s$^{-1}$,
is comparable to the $2\sigma$ upper limit on its spin-down power,
$\dot E < 7 \times 10^{33}$~ergs~s$^{-1}$.
Thus, it may be difficult to power the X-ray source with
spin-down energy.
Since the X-ray spectrum is consistent with thermal emission,
what about residual cooling of the neutron star?
While the X-ray luminosity of \psr\ is consistent with minimal
NS cooling curves for an age of $10^{3-4}$~yr \citep{pag04}, its
blackbody temperature implies an emitting area that is just $\approx
0.4\%$ of the NS surface.  Even taking into account that, in a real
NS atmosphere, a blackbody fit overestimates the effective
temperature and underestimates the area, the highly
modulated X-ray pulse must come from a small hot spot.
Although anisotropic conduction in a magnetized NS can enforce
temperature gradients on the surface, such models \citep{gep04,per06}
do not show spots as small and as hot as observed here.  Instead,
an external source of localized heating may be needed.

The open-field-line polar cap is the
most likely target area for external heating of the NS surface,
with the energy ultimately drawn
from rotation.  The canonical area of the polar cap is $A_{\rm pc} =
{2\pi^2 R^3 / {P c}} \approx 1 \times 10^{10}$~cm$^2$; this is
$\sim 14\%$ of the area implied by the blackbody fit to the X-ray spectrum.
Among theories of polar-cap heating, a maximum thermal luminosity
is predicted by \cite{wan98} from the outer-gap model for
$\gamma$-ray pulsars.  In this analysis, the X-ray luminosity of the
hot polar cap is limited
by the Goldreich-Julian $e^{\pm}$ current flow of $\dot N_0 \approx 2
\times 10^{31}\,(P/0.105\,{\rm s})^{-2}\,(B_p/10^{11}\,{\rm
G})$~s$^{-1}$ depositing an average energy per particle $E_f
\approx 4.3$~ergs.  The maximum luminosity is $L_{\rm bol} \approx f
E_f \dot N_0 < 4 \times 10^{31}$~ergs~s$^{-1}$.  Here, the fraction $f$
of the current reaching the surface is set to its estimated maximum
value of ${1\over 2}$.
Polar-cap heating models of \citet{har01,har02} predict even less X-ray
luminosity than \citet{wan98}.  Therefore, current theories fall
short of predicting the temperature and
luminosity of the X-ray emission from \psr\ in the context of
a spinning-down neutron star.

After the unexplained X-ray properties,
a lesser problem for \psr\ as an isolated pulsar is its
characteristic age $\tau_c \equiv P/2\dot P > 8$~Myr, which
exceeds by 3 orders of magnitude the SNR age, estimated dynamically
as $5.4-7.5$~kyr \citep{sun04}.  This would require the pulsar
to have been born at its current period to be associated with the SNR.
While not fundamentally contradicting the rotation-powered hypothesis,
this result is a definite example in support of a recent
population analysis that favors a wide distribution of radio
pulsar birth periods \citep{fau06}, even though \psr\ is not itself a
radio pulsar.  Furthermore, as magnetic field is generated by a
turbulent dynamo whose strength depends on the rotation rate
of the proto-neutron star \citep{tho93}, it is natural that
long-period pulsars would have the weaker $B$-fields at birth,
and the model of \citet{bon06} shows this.  \citet{bon06} also
find that a slowly rotating NS should have its $B$-field confined
to small-scale regions, with the global dipole that is responsible for
spin-down absent.

Finally, we note that there are no young radio pulsars known with
$B_p < 10^{11}$~G.  All such weakly magnetized radio pulsars are either
very old, with $\tau_c > 40$~Myr \citep{man05}\footnote{ATNF Pulsar Catalogue,
version 1.29, http://www.atnf.csiro.au/research/pulsar/psrcat/\hfil},
or are recycled, the result of accretion 
from a binary companion. \cite{fau06} conclude that there is no
evidence for magnetic field decay during the radio emitting life of
a pulsar, so the observered distribution of $B_p$ should resemble its
birth distribution.  Among the $B_p$ distribution of ordinary
(not recycled) radio pulsars,
\psr\ is certainly in the bottom 10\%, and maybe lower.  \psr\
may become a radio pulsar in the far future, but its weak $B$-field may
be preventing radio emission now by permitting accretion.  While
there is no evidence of a close binary companion in the timing of \psr\
(see Paper~1), it is possible that low-level accretion of SN debris
prevents it and other radio-quiet CCOs
from becoming radio pulsars for thousands or even millions of years.
We next consider whether the hypothesis of accretion from such a
fallback disk \citep[e.g.,][]{alp01,shi03,eks05}
better explains the X-ray timing and spectral
properties of \psr.

\subsection{Fallback Accretion?}

The fact that $\dot P$ is so small leads
us to review the viability of accretion of supernova debris
through a disk.  The X-ray luminosity of \psr\ can be powered
by accretion of $\dot m \approx 3 \times 10^{13}$ g~s$^{-1}$,
or only $\approx 0.1$ lunar masses of supernova debris over
the past 7.5~kyr.
The main barrier to disk accretion is the speed-of-light cylinder,
of radius $r_{\ell} = cP/2\pi = 5 \times 10^8$~cm.
If an accretion disk cannot penetrate the light cylinder, the NS
cannot interact with the disk, and it behaves as a isolated pulsar.
But if $B_p$ is as small as $10^{10}$~G, accretion at a rate
$\dot M \geq 10^{13}$~g~s$^{-1}$ can penetrate the light cylinder,
since the magnetospheric radius is then
\begin{equation}
r_m = 3.2 \times 10^8\,\mu_{28}^{4/7}\,\dot M^{-2/7}_{13}
\,\left({M \over M_{\odot}}\right)^{-1/7}\ {\rm cm},
\end{equation}
where the magnetic moment $\mu = B_p\,R^3/2 \approx 10^{28}B_{p,10}$ G~cm$^3$.
If so, the system is in the propeller regime,
in which matter flung out from
$r_m$ at a rate $\dot M$ takes angular momentum from the NS,
causing it to spin down.  The propeller spin-down rate is estimated as
\begin{equation}
\dot P \approx 2\,\dot M\,r_m^2\,I^{-1}\,P\,
\left(1- {P \over P_{\rm eq}}\right)
\end{equation}
\citep[e.g.,][]{men99,eks05}. Here $I \approx 10^{45}$ g~cm$^2$
is the NS moment of inertia, and
$P_{\rm eq} = 3.2\,\mu_{28}^{6/7}\,\dot M_{13}^{-3/7}\,
(M/M_{\odot})^{-5/7}$~s is the equilibrium, or minimum
period for disk accretion that is reached when $r_m \leq r_{\rm co}$.
The corotation radius is
$r_{\rm co} = [GM(P/2\pi)^2]^{1/3} = 3.7 \times 10^7$~cm.

While we have not measured a significant
$\dot P$ in order to determine the needed $\dot M$, we can adopt
the $2\sigma$ upper limit, $\dot P < 2.0 \times 10^{-16}$,
as an estimate of $\dot P$.
Combining equations (1) and (2), we have in the propeller scenario
\begin{displaymath}
\dot P \approx 2.2 \times 10^{-16}\,\mu_{28}^{8/7}\,\dot M_{13}^{3/7}\,
\left({M \over M_{\odot}}\right)^{-2/7}\,I_{45}^{-1}\,
\left({P \over 0.105\ {\rm s}}\right)\break
\end{displaymath}
\begin{equation}
\left(1- {P \over P_{\rm eq}}\right).
\end{equation}
We must distinguish here between $\dot M$, the matter expelled that is
responsible for the torque on the NS,
and $\dot m \approx 3 \times 10^{13}$~g~s$^{-1}$,
the matter accreted, which is responsible for the X-ray emission
from the surface, presumably at a magnetic pole of the neutron star.
For the propeller model to be self-consistent,
$\dot M$ must be  $> \dot m$, which is possible according to
equation (3) as long as $B_p < 10^{10}$~G.  Even so,
$\dot M$ does not contribute significantly to the X-ray luminosity because
of the much weaker gravitational potential at the
magnetospheric radius $r_m$ from which it is expelled.
If $B_p$ is as small as $7 \times 10^8$~G, then \psr\ is
not a propeller, but a ``slow rotator,''
with $r_m \leq r_{\rm co}$ and $P_{\rm eq} \leq P$.
In this spin-up regime, it would be even more difficult to
measure $\dot P$, since
\begin{equation}
\dot P \approx -1.3 \times 10^{-18}\,\mu_{27}^{2/7}\,\dot m_{13}^{6/7}\,
\left({M \over M_{\odot}}\right)^{3/7}\,
\left(P \over 0.105\ {\rm s}\right)^2
\end{equation}
\citep{gos79}.

These equations also apply during the prior evolution of the pulsar
and show that, even if the accretion rate was orders of
magnitude higher in the past,
the spin-up and spin-down times $P/\dot P$ are much longer than the age
of the SNR.  There has not been enough time for propeller accretion
to have spun the pulsar down from a much smaller $P$, so even in
this model it was born at essentially its present $P$.

Thus, we regard either propeller spindown or accretion in a
weak magnetic field as consistent with
the X-ray spectral and timing properties of \psr.
If accreting, its emitting region could be larger than
the open-field-line polar cap because it would be connected
to magnetic field lines extending to $r_m$, which is $<r_{\ell}$.
Neither equation (3) nor equation (4) requires there to be much
torque noise at the level that present data can measure.
While detected flickering is
an indicator of accretion in X-ray binaries even in quiescence
\citep{rut07}, we do not have strong evidence of variability
of \psr.  But it is not clear that the processes in binaries can
be extrapolated to accretion at such low rates from a fossil disk.
Finally, because of the large distance and interstellar extinction to
\psr, as well as its crowded optical field, existing optical data do
not rule out the presence of a fallback accretion disk around \psr\
(see Paper~1 for details).

\subsection{Comparison with 1E 1207.4$-$5209 and Other CCOs}

Difficulties in understanding the timing behavior of
another CCO, the 424-ms pulsar \one\
\citep{zav04,woo06}, were resolved recently
by \citet{got07}, who showed that it is a steady
rotator with $\dot E < 1.5 \times 10^{32}$ ergs~s$^{-1}$
and $B_p < 3.5 \times 10^{11}$~G.  This makes \one\
very similar to \psr\ in its physical parameters.
Accretion from a fossil disk of supernova debris
\citep{alp01,shi03,eks05,liu06} was one of the
theories considered by \citet{zav04} to explain
the now defunct timing irregularities of \one. 
But accretion may still be needed to explain the
radio-quiet nature of \one, 
and the details of its X-ray spectrum.  However,
unlike \psr, upper limits on optical/IR emission from \one\
are comparable to what is expected from a geometrically
thin, optically thick disk accreting at the rate required to
account for its X-ray luminosity \citep{zav04,wan07}.
Therefore, it may prove necessary to invoke a
radiatively inefficient flow, or perhaps even accretion of
solid particles \citep{cor06}, in order to proceed
with this theory for \one.
We propose that this may be
the first phase in the life of those neutron stars born
rotating slowly, and with weak magnetic fields.

A unique phenomenon displayed by \one\ is the set of
broad absorption lines in its soft X-ray spectrum.
They are centered at $0.7$~keV and $1.4$~keV \citep{san02,mer02},
and possibly at 2.1~keV and 2.8~keV \citep{big03,del04}, although the
reality of the two higher-energy features has been
disputed \citep{mor05}.
Proposed absorption mechanisms include electron cyclotron
in a weak ($8 \times 10^{10}$~G) magnetic field \citep{big03,del04},
atomic features from singly ionized helium in a strong
($2 \times 10^{14}$~G) field \citep{san02,pav05},
and oxygen or neon in a normal ($10^{12}$~G)
field \citep{hai02,mor06}.  A detailed critique of these
models is beyond the scope of this paper.  We only remark that:
(1) local values of $B$ measured from X-ray spectral features may
differ from the global dipole $B$ that is measured by timing,
especially if the dipole component is weak \citep{bon06}, and (2)
the attractive feature of the electron cyclotron model for \one,
the equal energy spacing of its lines, is in harmony with
the present work on \psr\ which, by different arguments,
requires a weak surface magnetic field.

\citet{xu03} argued that electron cyclotron harmonics
of roughly equal depth could be produced because,
even though the oscillator strength of the
harmonic is much less than that of the fundamental, resonant
cyclotron radiative transfer could equalize the line strengths.
\citet{liu06} also explained how objections to the cyclotron model
involving the relative oscillator strengths
can be overcome by assuming that the
fundamental is optically thick while the harmonic is optically
thin, resulting in comparable equivalent widths.
They also proposed a specific geometry that accounts
for the strength of the lines, which also accommodates the
small pulsed fraction observed from \one.
This involves a column rising along the magnetic axis and
viewed from the side.  Finally, \citet{liu06} remarked that such a
column may be more easily maintained in an accretion scenario.

While the lower-quality spectra of \psr\ do not
yet show evidence of absorption features, cyclotron lines could
be weak if either the column density in which they are formed
or the viewing geometry is not favorable, or especially if the magnetic
field is so small as to shift them below the observable X-ray band.
The fitted interstellar column density to \psr,
$\nh = 1.4 \times 10^{22}$~cm$^{-2}$, is much larger than that
to \one, $\nh = (3-10) \times 10^{20}$~cm$^{-2}$ \citep{mer02}.
Therefore, the soft X-ray region in which the strongest lines
in \one\ are seen is partly suppressed in the case of \psr.
Finally, it is possible that the X-ray spectrum of \psr\ is not a pure
blackbody as fitted, but is affected by unmodelled,
broad cyclotron absorption at the low-energy end.

The half dozen CCOs \citep[for a review, see][]{pav04} are similar
in their X-ray luminosities and temperatures.  Therefore,
they may be compatible with the model that
we propose for \psr\ and \one.  The only piece of
evidence that sharply contradicts such unification
is the apparent historical outburst of the CCO in the
Cas~A SNR that was inferred from infrared light echos
detected with the {\it Spitzer Space Telescope} \citep{kra05}.
The geometry and sharpness of the echos implies that beamed emission
from the central source flared for less than a few weeks about
50 years ago, and the radiation reprocessed into thermal emission
from dust require an equivalent isotropic flare of
$\sim 2 \times 10^{46}$~ergs.  Because such energetic
outbursts are seen on similar timescales from soft gamma-ray
repeaters, \citet{kra05} proposed that the Cas~A CCO is a
quiescent magnetar, at the other extreme of
magnetic field strength from \psr.  A convincing determination
of the magnetic field strengths of the Cas~A CCO and others
will have to await discovery of their pulsations.

\subsection{Application to SN 1987A}

It has long been recognized that the non-detection of the expected
pulsar in the remnant of SN~1987A can be explained if the
NS was born with a weak magnetic field or a long rotation
period \citep{oge04}, and that several CCOs,
including the one in Cas~A, would be undetected if placed in
SN~1987A \citep{gra05}.  However, it was not established
until now that a CCO can have essentially the same spin-down
luminosity at birth that it has at an age of 400 or $10^4$~yr,
i.e., that it is born with both a weak $B$-field and a long $P$.
This new result means that a common type of neutron
star can emit less than the observed limits from SN~1987A 
even if 100\% of its spin-down power is reprocessed into
IR emission by dust in the surrounding
SN ejecta.  The relevant luminosity limits to be satisfied
for a point source inside the ring of SN~1987A
are $L_x(2-10\,{\rm keV}) < 1.5 \times 10^{34}$ ergs~s$^{-1}$
corrected for extinction
\citep{par04}, $L(2900-9650$~\AA) $< 8 \times 10^{33}$ ergs~s$^{-1}$
corrected for dust absorption \citep{gra05},
and $L(10\,\mu{\rm m}) < 1.5 \times 10^{36}$ ergs~s$^{-1}$ for
dust emitting at $T \approx 90-100$~K \citep{bou04}.  This mid-IR
luminosity can be accounted for by radioactive decay of $^{44}$Ti,
and therefore represents a very conservative upper limit on the
spin-down power of an embedded pulsar.
At an age of $10-20$~yr, a cooling NS need
emit only $\approx 3 \times 10^{34}$ ergs~s$^{-1}$ of
soft X-rays at a temperature of $2.5 \times 10^6$~K \citep{yak02},
some of which is absorbed by SN ejecta or ISM in the LMC.
The upper limit on spin-down power from \psr,
$\dot E < 7 \times 10^{33}$ ergs~s$^{-1}$,
is a challenge to detect in SN~1987A,
and $\dot E < 1.5 \times 10^{32}$ ergs~s$^{-1}$
from \one\ would be impossible.  So we conclude that
a CCO is a promising model for the unseen NS in SN~1987A.

\section{Conclusions and Future Work}

We obtained three new X-ray timing points on \psr\ that, in combination
with two previous observations, reveal no significant change in spin
period on time scales ranging from 1 week to 2.4 yr.  In the dipole
spin-down formalism, this implies $\dot E < 7 \times 10^{33}$ ergs~s$^{-1}$
and $B_p < 1.5 \times 10^{11}$~G.
Such a low $B$-field is not seen in young
radio pulsars, and while it does not demand a special explanation,
it does favor accretion as a source of the small,
hot thermal region that is responsible for the highly modulated
X-ray pulsations,
which may exceed the spin-down power of the isolated NS,
and strains any plausible interior cooling or external heating model.

We showed that the observered spin period, 105 ms, in combination with
a weak field, $B_p \sim 10^{10}$~G, would allow disk
accretion in the propeller regime, which may be a more satisfactory
model overall.  If $B_p$ is even weaker, $< 10^9$~G, \psr\ could
be a ``slow'' accretor.
Considering also the properties of the CCO \one,
we speculated that this
class may have weak magnetic fields as a result of slow
natal rotation, which enables them to accrete. While we do
not decisively favor accretion over a rotation-powered pulsar,
we emphasize that for \psr\ a weak magnetic field is an unavoidable
feature of either interpretation.  Such low-$B$ births may
be very common if they also occurred in Cas~A and/or SN~1987A.

A radio pulsar detection of \psr\ would demonstrate cleanly
that it is rotation powered, but our negative results
so far are inconclusive because there are several reasons
why a rotation-powered pulsar may not be detected in radio.
Continued X-ray monitoring will provide a more sensitive
test for spin-down or accretion torques, perhaps eventually
in a definitive manner, and more evidence of
whether or not the X-ray flux and pulse profile are variable.
It is now important to obtain a phase-connected
timing solution, the most efficient and sensitive method of measuring
a small $\dot P$.   If obtained soon, a well-timed series of
observations should allow a coherent solution to be fitted
retroactively back to 2004.  Also, considering the possibility 
that a weak magnetic field and accretion may be responsible for the
absorption lines in the soft X-ray spectrum of \one, a deeper X-ray
spectral study of \psr\ may be revealing.  Finally, in order to
test the general applicability of these ideas to the class of CCOs,
more sensitives searches for their pulsations are required.

\acknowledgments

This investigation is based on observations obtained with \xmm, an ESA
science mission with instruments and contributions directly funded by
ESA Member States and NASA.  Support for this work was provided by
NASA through {\it XMM\/} grant NNX06AH95G and {\it Chandra} Award
SAO GO6-7048X issued by the {\it Chandra} X-ray Observatory Center,
which is operated by the Smithsonian Astrophysical Observatory
for and on behalf of NASA under contract NAS8-03060.
The GBT is
operated by the National Radio Astronomy Observatory, a facility of the
National Science Foundation operated under cooperative agreement by
Associated Universities, Inc.

\end{document}